\documentclass[jkps,preprint,fleqn,showpacs,showkeys]{revtex4}
\usepackage{graphicx}
\usepackage{amssymb}
\usepackage{amsmath}
\usepackage{bm}
\begin{document}
\setcounter{page}{0}

\title[]{The design of the optimized muon beamline}

\author{Suyong \surname{Choi}}

\author{Jeongwon \surname{Park}}

\author{Youn Jung \surname{Roh}}
\email{kuyoun@korea.ac.kr}
\thanks{Fax: +82-2-3290-3968}

\affiliation{Department of Physics, Korea University, Seoul 136-713}

\begin{abstract}
We designed an optimized antimuon beamline with the simulation using G4beamline and TRANSPORT software packages for Heavy Ion Accelerator Project in South Korea. This research suggests the optimized muon beamline can transport $2.4\times10^{8}$ antimuons per second to a circle with a radius of 3 cm. In terms of muon rate, this is competitive with world leading muon beam facilities.
\end{abstract}


\maketitle

\section{INTRODUCTION}

The Heavy Ion Accelerator Project's RAON affiliated to the Institute for Basic Science(IBS) is a Korean heavy ion accelerator that will be constructed in Daejeon in 2017 \cite{BaseLine}. The name RAON is a pure Korean term that means ``Pleasant". It basically aims to examine the origin of the strong force by studying the interactions between the nuclei that construct rare isotope and is expected to train a lot of  domestic experts of accelerator\cite{RAON}.

 As a branch of the RAON accelerator, we propose a construction of the first muon beamline in Korea. The muon beam is very difficult to handle because they start life as a hot gas \cite{Muon_Cooling}, but the scientists can perform the various research with the muon beam. Paul Scherrer Institute (PSI) \cite{PSI} and Japan Proton Accelerator Research Complex (J-PARC) \cite{J-PARC} overcame this problem and run their own muon beamlines, but Korea does not have one yet. If Korea has the muon beamline, it will enhance the level of basic science of Korea remarkably and raise the global competitiveness in science area.

\begin{figure}[!h]
\noindent
\centering
\vspace{-2mm}
\includegraphics[width=1\textwidth]{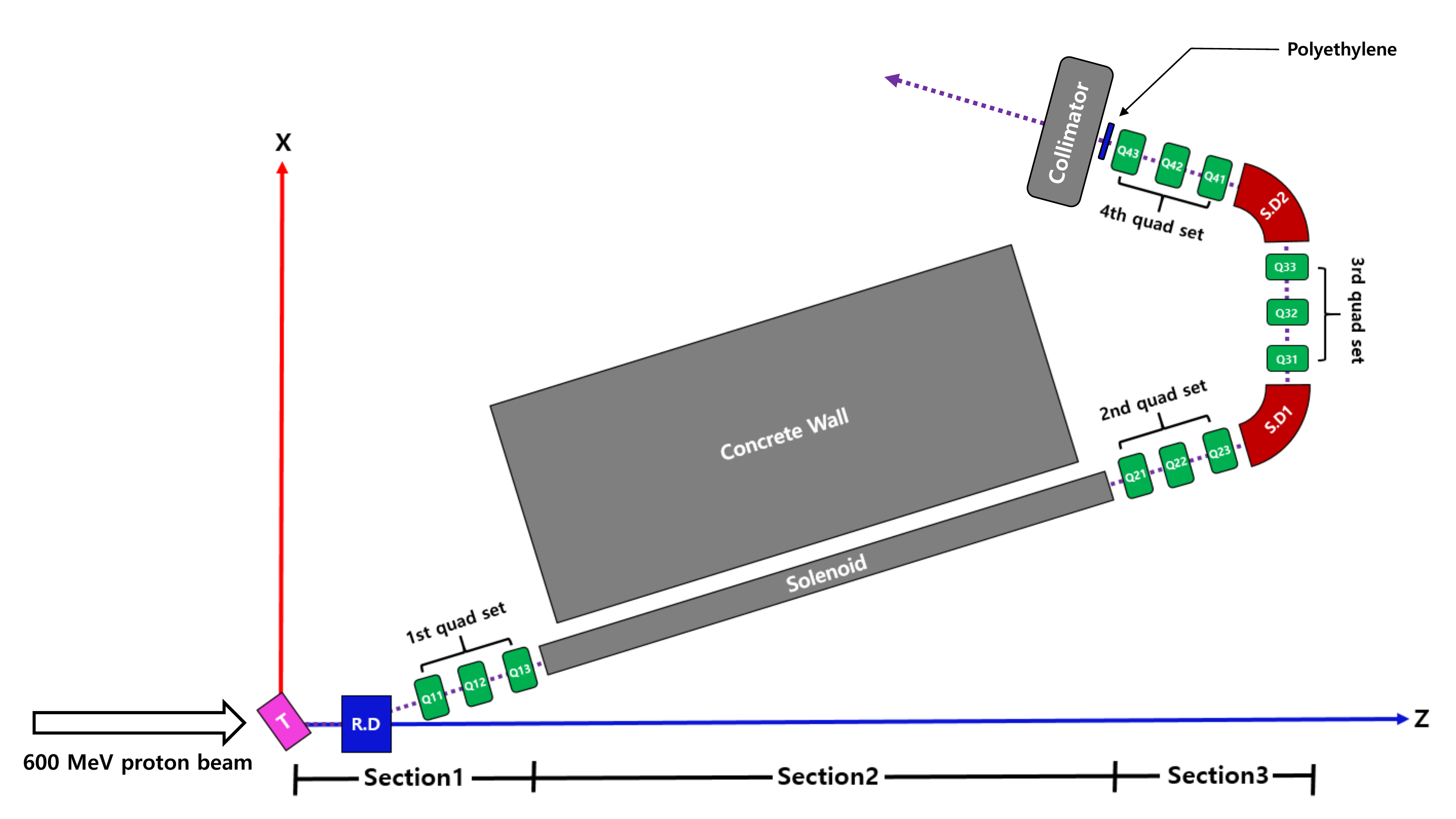}
\caption{\label{Top}The schematic view of the optimized muon beamline. It is divided into three sections : $\pi^+$ collecting section (Section1), $\pi^+$ decay section (Section2) and $\mu^+$ collecting and purification section (Section3). The labels T, R.D, Q and S.D mean the Target, Rectangular Dipole, Quadrupole and Sector Dipole, respectively, and the dashed line means the design orbit on which ideally all particles should move. This beamline is designed to be 20 m wide and 26 m long. The concrete wall is used to block the particles that fly from the target to the collimator.}
\par
\end{figure}

Stimulated by this intention, we have designed muon beamline as shown in FIG. \ref{Top}. This beamline is designed with 600 MeV kinetic energy proton beam on a fixed target, and the beamline elements are optimized to provide 130 MeV/c antimuons with low contamination. In order to perform this study, we used special software packages, G4beamline \cite{G4beamline}: the particle tracking simulation program based on Geant4 \cite{Geant4} and Transport \cite{Transport}: the calculation program that gives the appropriate values to be applied to the magnet elements in the beamline to simulate particle production and decays, and their transport through beamline optics.

\section{Simulation}

The simulations were conducted with the G4beamline Monte Carlo package. For applications of muon beamline in low energy physics, G4eamline recommends to use the QGSP$\_$BERT physics list which uses the quark gluon string model and Geant4 Bertini cascade \cite{QGSP_BERT}. Each run consisted of $10^{7}$ proton events on the fixed target and the events were generated by pseudo random number generator with time.

\subsection{$\pi^{+}$ Collecting Section}
\begin{figure}[!h]
\noindent
\centering

\includegraphics[width=1\textwidth]{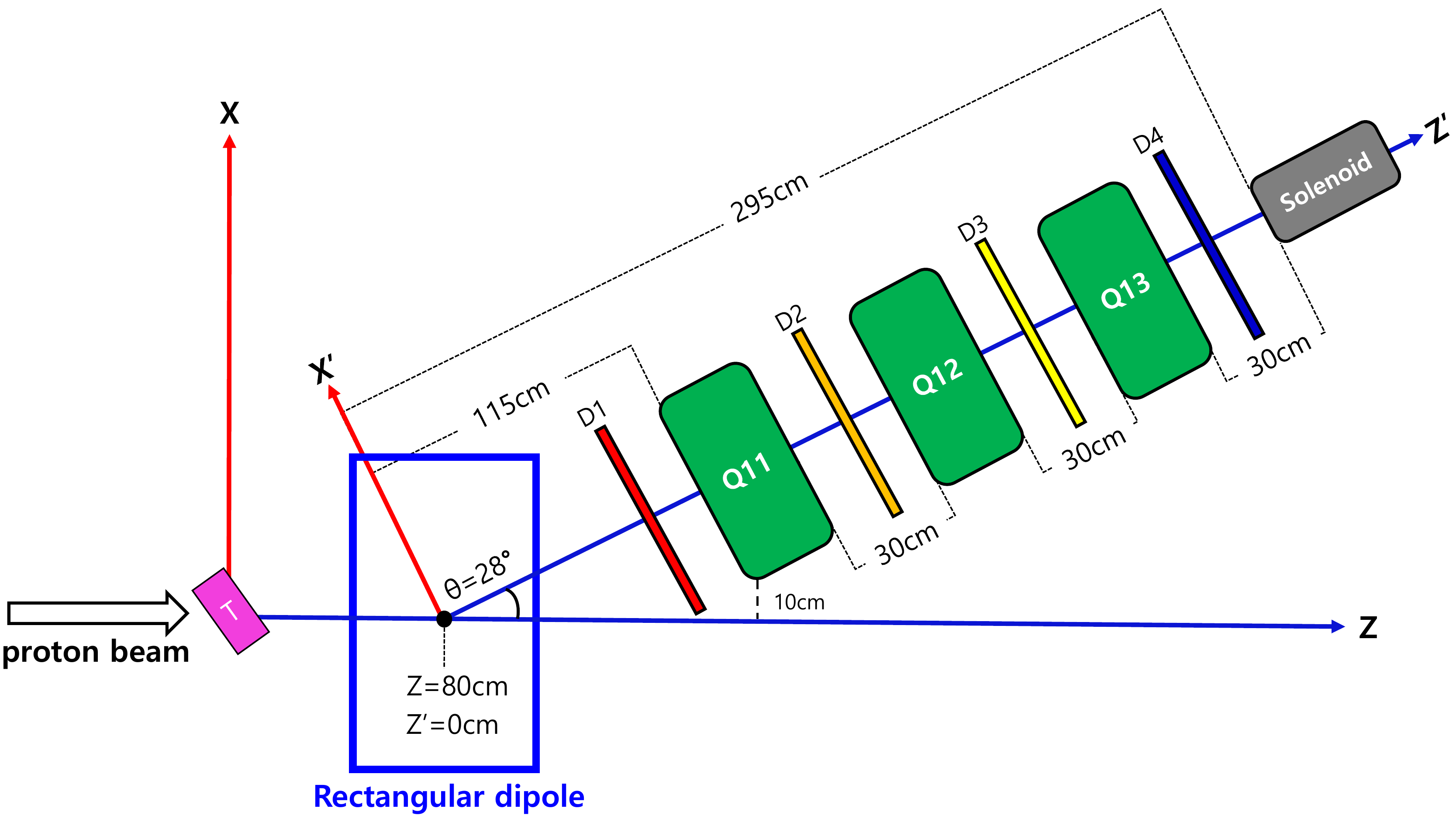}\\
\caption{\label{1st_Quad_coordinate}The schematic view of $\pi^+$ collecting section. Label D means the imaginary detectors to monitor beam profile.}
\par
\end{figure}

\begin{figure}[!h]
\noindent
\centering
\vspace{-4mm}
\includegraphics[width=0.5\textwidth]{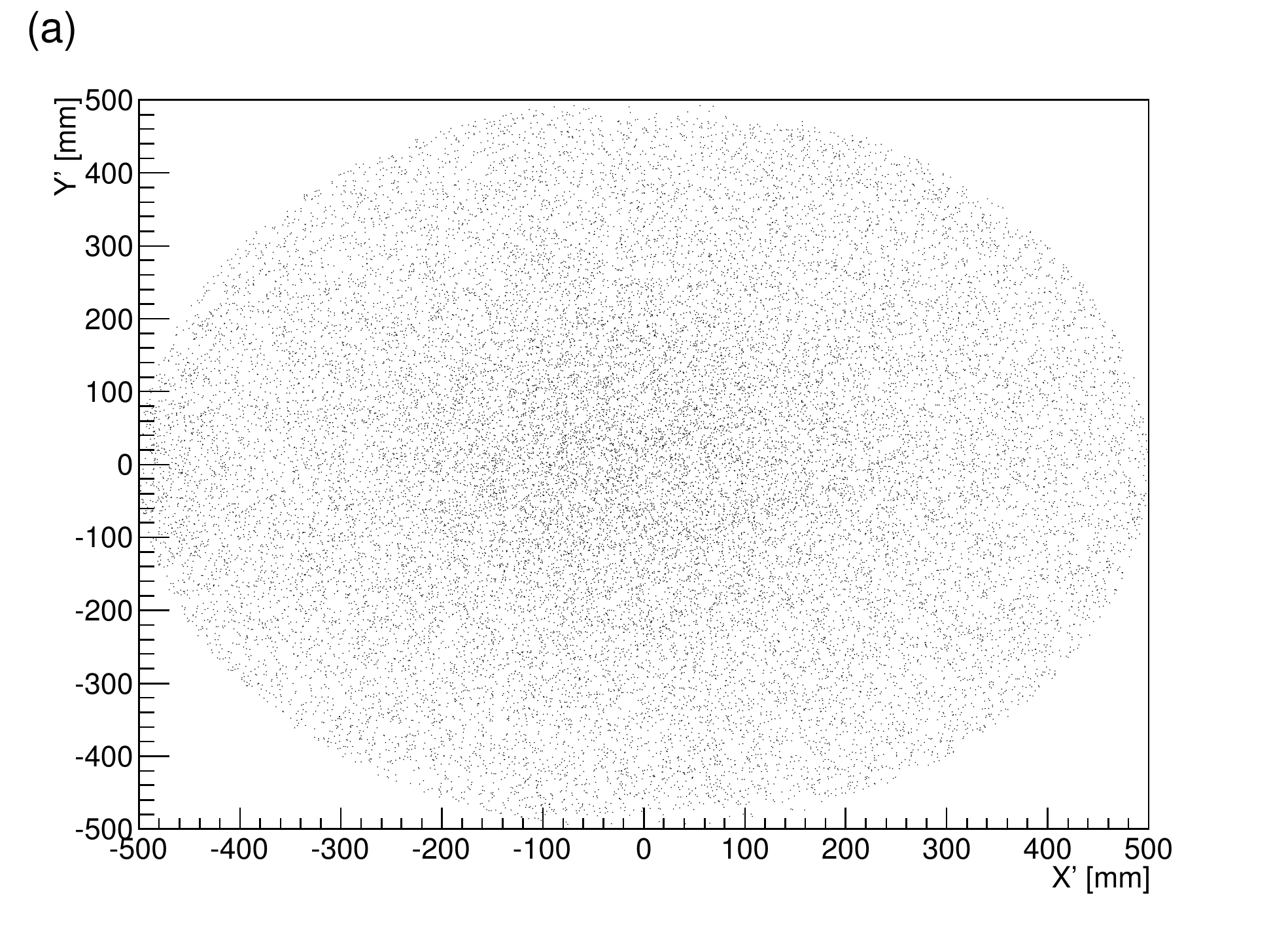}
\hspace{-0.5cm}
\includegraphics[width=0.5\textwidth]{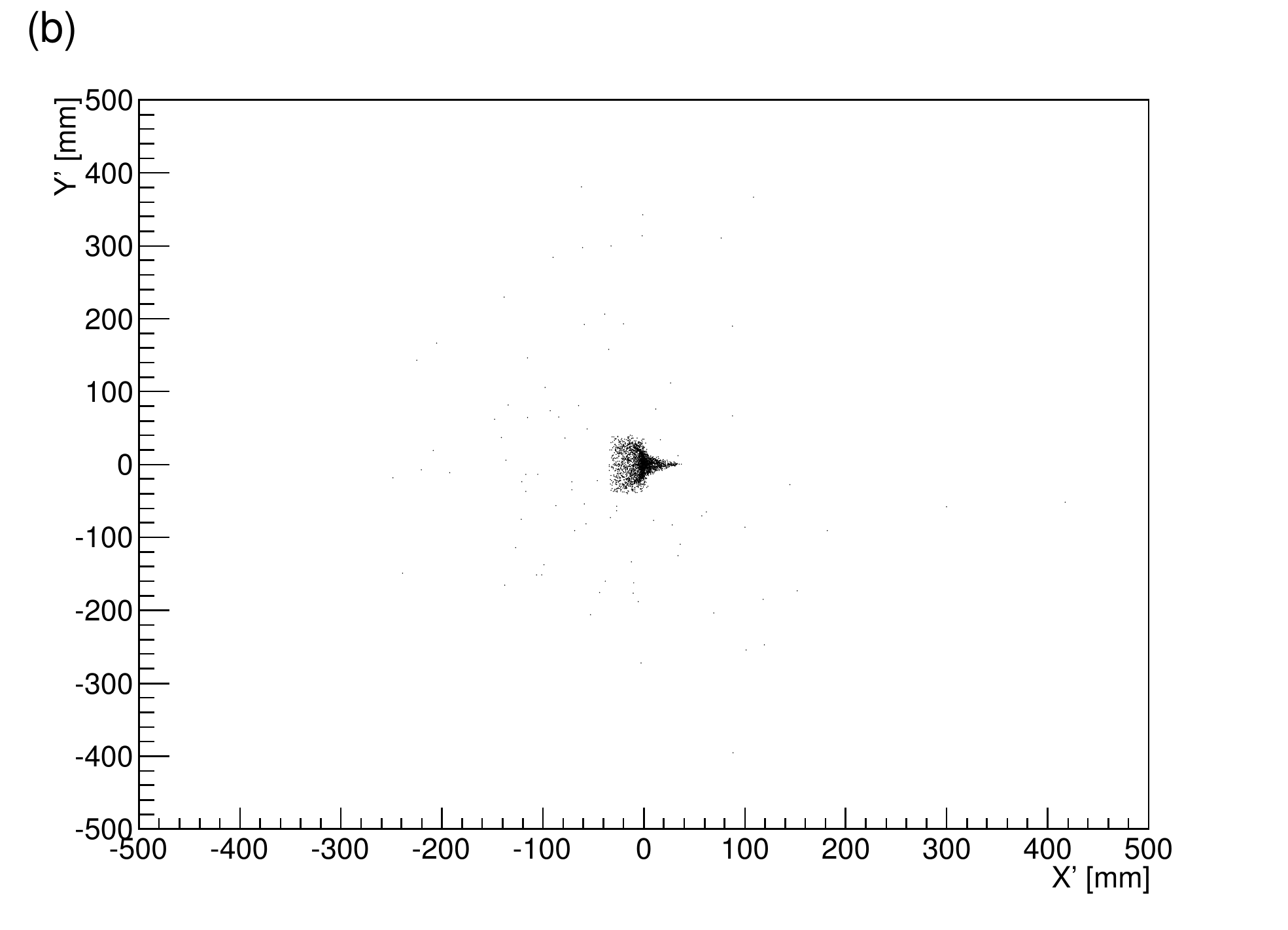}\\
\caption{\label{1st_Quad_Det}The shape of 190 MeV/c$\sim$210 MeV/c $\pi^{+}$ ($4\times10^{8}$ protons on the fixed target). (a) and (b) show the information of the D1 and D2, respectively. It can be shown from (b) that most pions exist in the region of a circle with a radius of 10 cm (the inner radius of the solenoid).}
\par
\end{figure}

\hspace{1.15cm}
We consulted J-PARC and placed 20 mm graphite target \cite{J_Target}, and from the view point of the beam sharing with the surface muon study group, we decided to rotate the target $45^{\rm \circ}$ with respect to the $\hat{z}$ direction \cite{Sorimsa}. The nuclear reactions between the 600 MeV kinetic energy proton beam and the nucleus of carbon produce various secondary particles and the 1219 MeV/c protons play the most significant role for the contamination. We choose 200 MeV/c $\pi^+$ and separated them from the protons with the rectangular dipole that applies 0.5 T in the -$\hat{y}$ direction. If we assume both the protons and $\pi^+$ move on the design orbit, 1219 MeV/c protons draw a circle of 8.13 m radius while about 200 MeV/c pions draw a circle of 1.3 m radius. As a result, both particles proceed into different directions and we can separate the $\pi^+$ from the protons. Then, the first quadrupole set is used to gather and focus the bent pions. Each of the quadrupoles is of cylindrical with a inner radius of 20 cm, a outer radius of 50 cm, and 30 cm length and are located on the new design orbit that is rotated 28$^{\circ}$ with respect to $\hat{z}$ direction as shown in FIG. \ref{1st_Quad_coordinate}. The field gradients of the first quadrupole set to focus 200 MeV/c pions are given 3.88 T/m, -5.20 T/m and 3.88 T/m, respectively and this triplet successfully focuses the pions to the center as shown in FIG. \ref{1st_Quad_Det}.

\subsection{$\pi^{+}$ Decay Section}
\begin{figure}[!h]
\noindent
\centering
\includegraphics[width=0.5\textwidth]{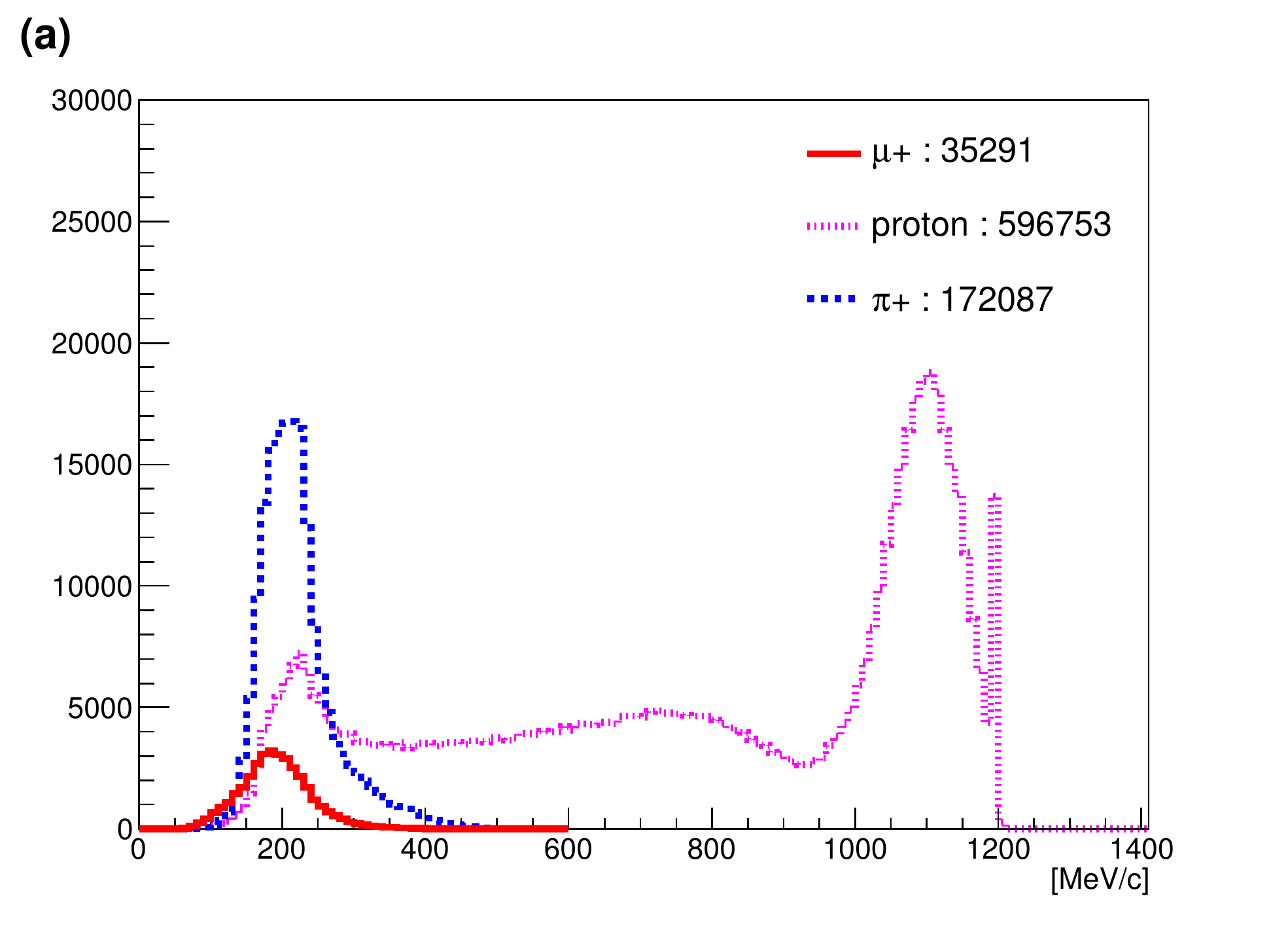}
\hspace{-5mm}
\includegraphics[width=0.5\textwidth]{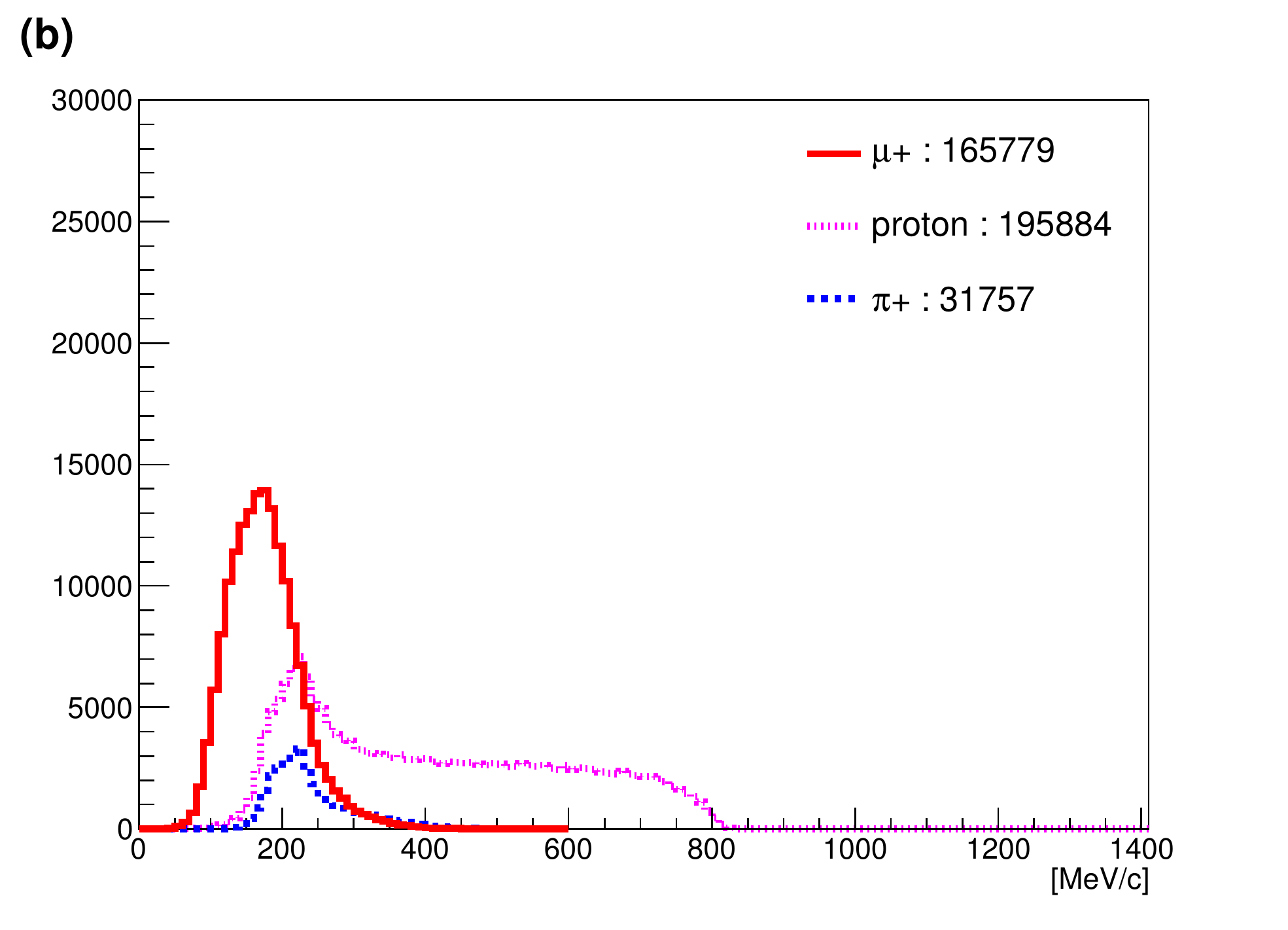}
\vspace{-2mm}
\caption{\label{Pion_vs_Muon_Bf_Af_Sol}The momentum distributions of the protons, pions and the antimuons. (a): Entering the solenoid. (b): Exiting the solenoid($5\times10^{9}$ protons on the fixed target).}
\par
\end{figure}

\hspace{1.15cm}
For $\pi^+$ decay section, we use a solenoid with a inner radius of 10 cm, a outer radius of 17 cm outer radius, 20 m length, and 5 T magnetic field. This is used to take the pions to the next element providing them the enough time to decay into antimuons without suffering collision. FIG. \ref{Pion_vs_Muon_Bf_Af_Sol} shows the momentum distribution of the protons, pions and the antimuons entering and exiting the solenoid. The number of pions dominates over the number of antimuons before the solenoid, but the situation is reversed afterwards.

\begin{figure}[!h]
\noindent
\centering
\vspace{-4mm}
\includegraphics[width=0.5\textwidth]{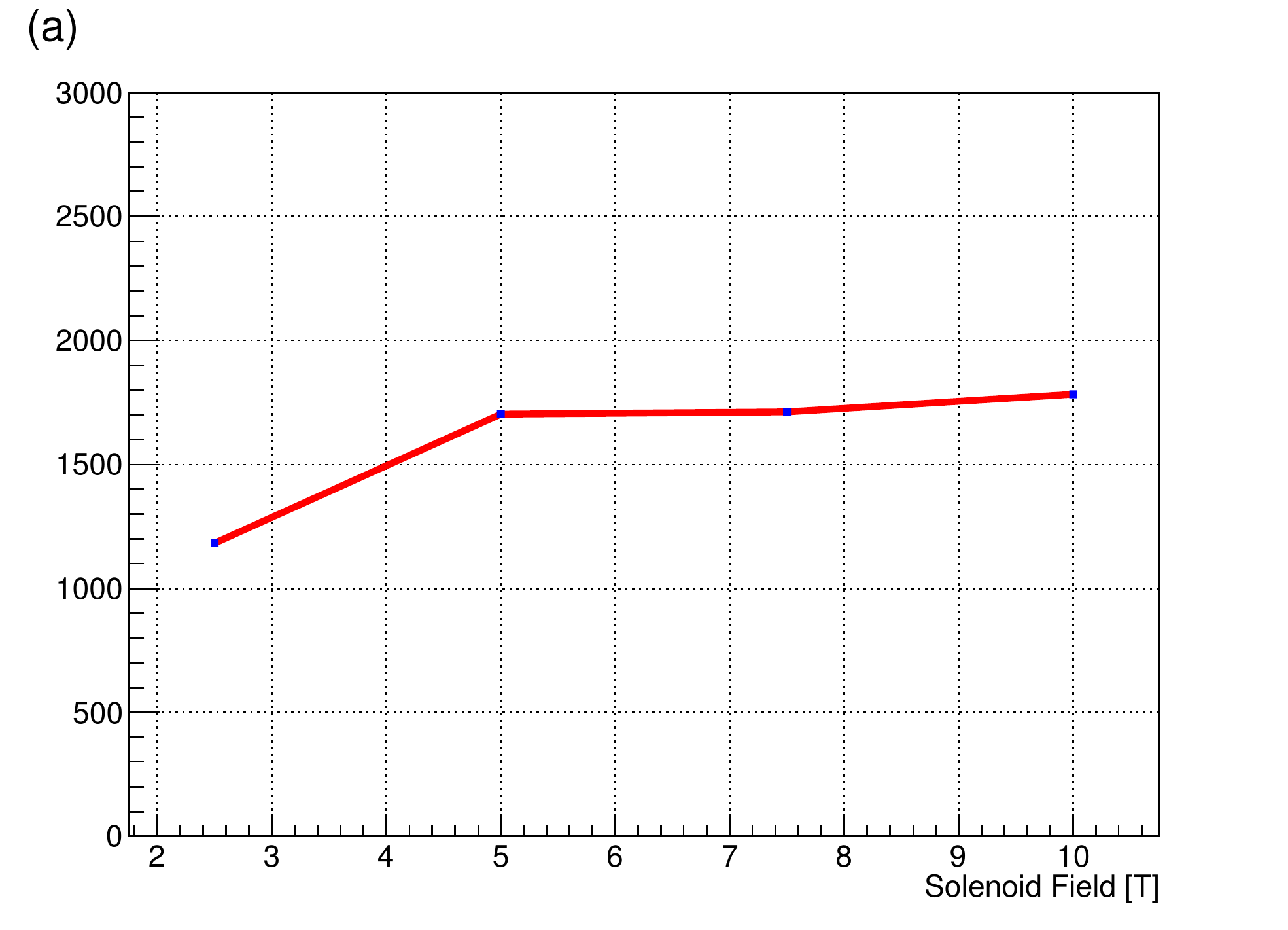}
\hspace{-0.5cm}
\includegraphics[width=0.5\textwidth]{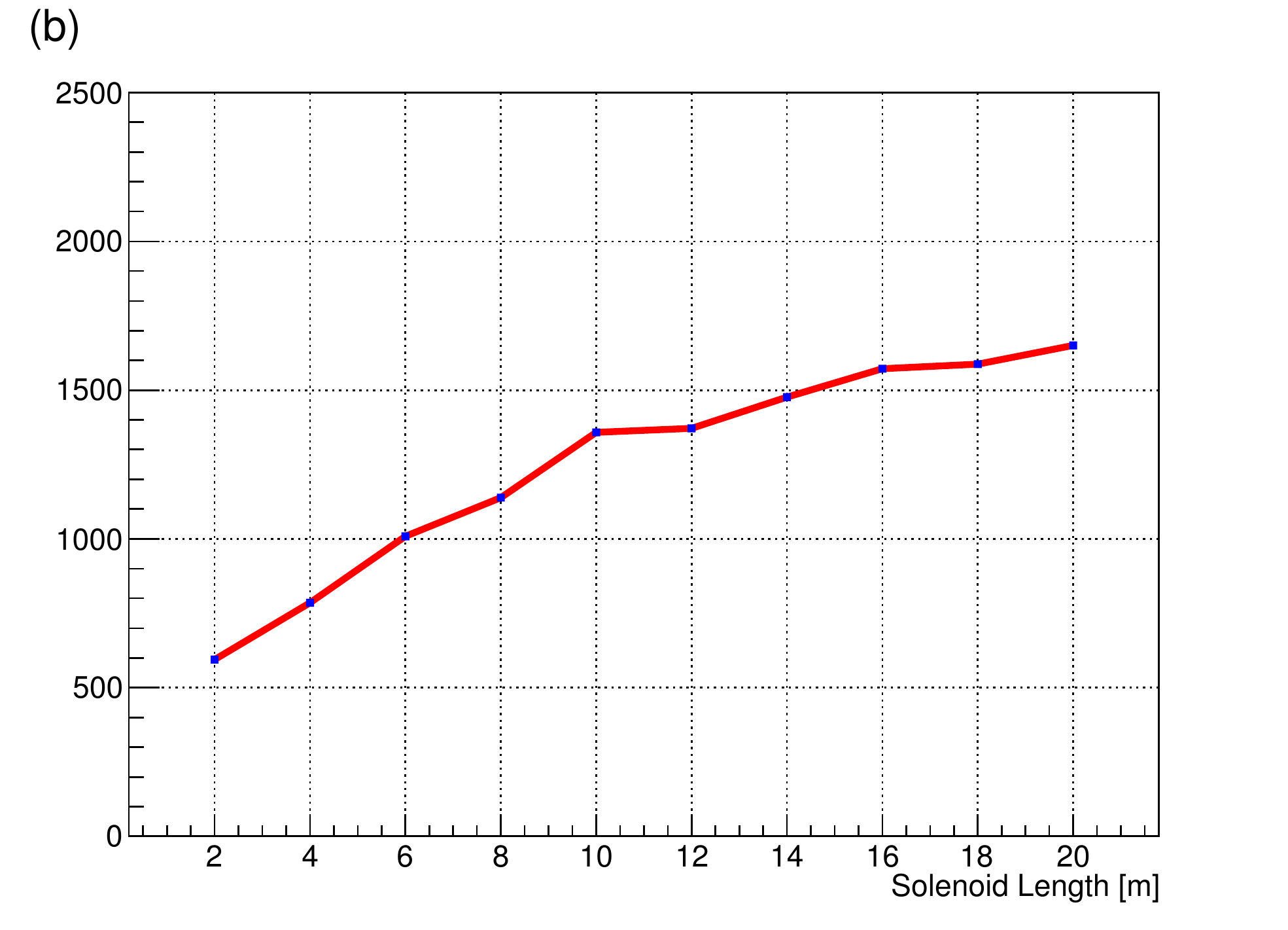}
\vspace{3mm}
\caption{\label{Sol_Change_B_L}The number of antimuons exiting the solenoid ($5\times10^{7}$ protons on the fixed target). (a): Different fields with 20 m solenoid. (b): Different lengths with 5 T solenoid.}
\par
\end{figure}

Generally, the stronger magnetic field of the solenoid, larger the number of antimuons as shown in FIG. \ref{Sol_Change_B_L}-(a). This is because the stronger magnetic field helps the charged particles to have smaller radius of helical motion, so that they can avoid the collision with the solenoid body. However, it does not increase drastically after 5 T, so we determined to apply this field to the solenoid, similar to $\mu$E1 beamline at PSI \cite{PSI}.

In addition, the longer length of the solenoid gives the larger number of antimuons as shown in FIG. \ref{Sol_Change_B_L}-(b) because the longer length provides the pions the increased probabilities to decay into antimuons without collision. TABLE \ref{tab:Soleng_Beam_Info} shows the quality of the antimuon beam exiting the solenoid does not depend significantly on the length of the solenoid as long as the field of the solenoid is kept at 5 T. This means that the optimized values of the elements after the solenoid can be kept constant regardless the length of the solenoid. Hence, we chose 20 m and optimized the magnets after the solenoid because this length gives a lot of antimuons to optimize the magnets and can be altered without the modifications of the optimized values after the solenoid.

\vspace{-0.1cm}

\begin{table}[!h]
\centering
\begin{tabular}{|c|c|c|c|c|c|c|}
\hline
{ } & { } & {4 m}& {8 m}& {12 m}& {16 m}& {20 m}\\
\hline
\hline
{~\tt x$\rm~[mm]$}& {\tt~~mean~~}        &{~\tt -1.8~}    &{~\tt -4.8~}     &{~\tt -5.3~}     &{~\tt -5.8~}    &{~~\tt -5.5~~}\\
\hline
{ }& {\tt RMS}                    &{~\tt 33.0~}    &{~\tt 35.5~}     &{~\tt 34.3~}     &{~\tt 34.4~}    &{~~\tt 34.1~~}\\
\hline
{~\tt y$\rm~[mm]$}& {\tt mean}        &{~\tt 3.9~}     &{~\tt 2.1~}      &{~\tt 4.4~}      &{~\tt 2.9~}     &{~~\tt 2.9~~}\\
\hline
{ }& {\tt RMS}                    &{~\tt 33.8~}    &{~\tt 33.3~}     &{~\tt 33~}       &{~\tt 33.0~}    &{~~\tt 33.2~~}\\
\hline
{~~\tt px/pz$\rm~[rad]$~~}& {\tt mean}   &{~\tt -0.003~}  &{~\tt -0.017~}   &{~\tt -0.004~}   &{~\tt -0.012~}  &{~~\tt -0.003~~}\\
\hline
{ }& {\tt RMS}                    &{~\tt 0.15~}    &{~\tt 0.15~}     &{~\tt 0.15~}     &{~\tt 0.14~}    &{~\tt 0.15~}\\
\hline
{~~\tt py/pz$\rm~[rad]$~~}& {\tt mean}   &{~\tt -0.010~}  &{~\tt -0.005~}   &{~\tt 0.007~}    &{~\tt -0.004~}  &{~~\tt -0.003~~}\\
\hline
{ }& {\tt RMS}                    &{~\tt 0.16~}    &{~\tt 0.17~}     &{~\tt 0.15~}     &{~\tt 0.15~}    &{~~\tt 0.15~~}\\
\hline
\end{tabular}
\caption{\label{tab:Soleng_Beam_Info}The lengths dependence of the antimuon beam exiting the solenoid whose field is 5 T ($5\times10^{7}$ protons on the fixed target).  As long as the field is kept constant, the quality changes much slightly.}
\end{table}

\subsection{$\mu^{+}$ Collection and Purification Section}
\begin{figure}[!h]
\noindent
\centering
\includegraphics[width=0.9\textwidth]{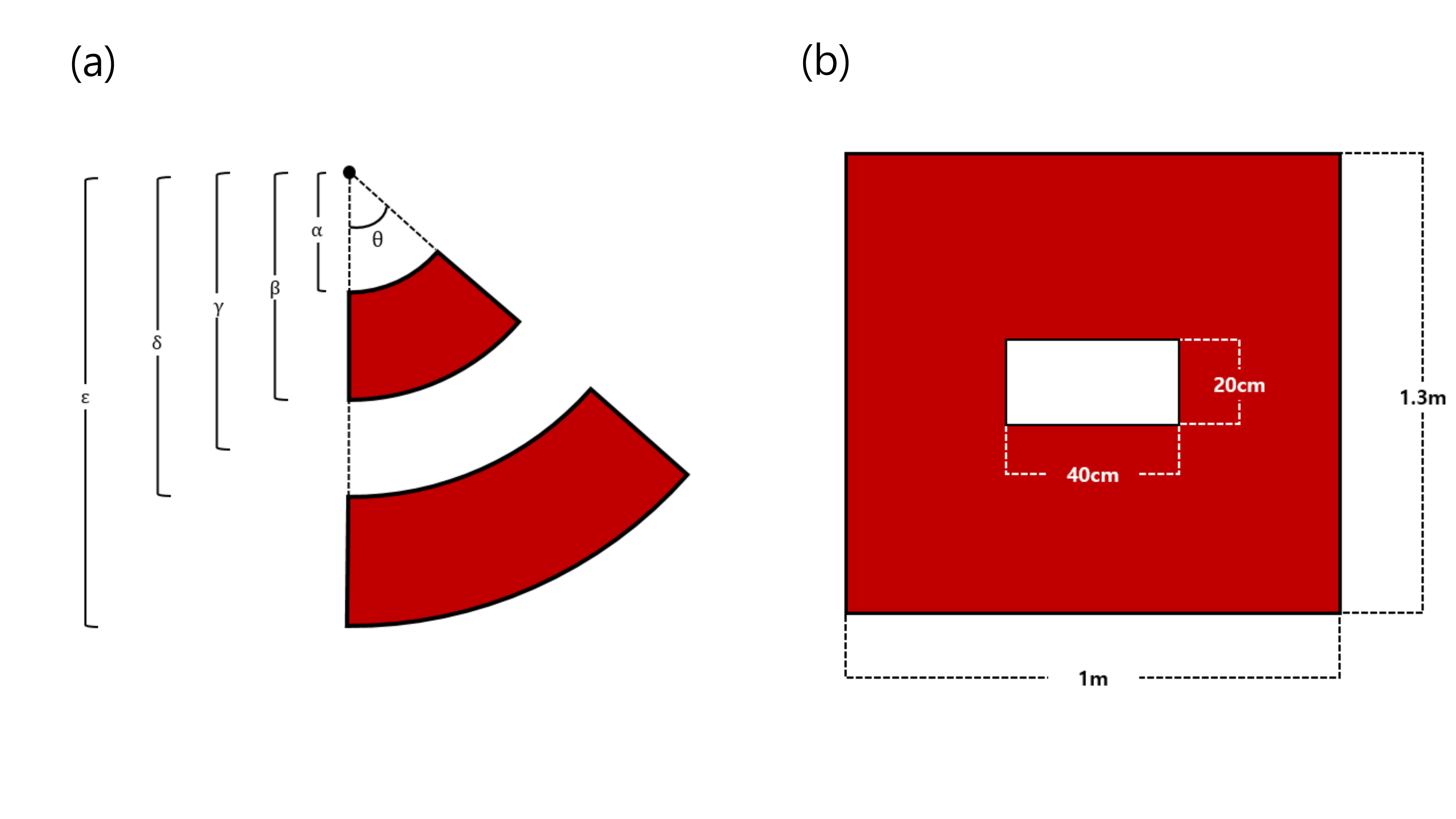}
\vspace{-8mm}
\caption{\label{1st_Sector}The schematic view of the sector dipole. (a): Top view. $\alpha$=1.17 m, $\beta$=1.47 m, $\gamma$=1.67 m, $\delta$=1.87 m and $\epsilon$=2.17 m. (b): The front view.}
\par
\end{figure}

\hspace{1.15cm}
FIG. \ref{Pion_vs_Muon_Bf_Af_Sol}-(b) shows that the protons and $\pi^+$'s act as dominant contamination after the solenoid. Since they are of higher momentum than antimuons, we applied the momentum selection with the dipole that is optimized to bend the antimuons whose momentum range is about 130 MeV/c. Especially, we used a sector dipole in the antimuon collection section instead of the rectangular dipole because we do not have to cover the proton beamline any more. The schematic view of the sector dipole is shown in FIG. \ref{1st_Sector}.

\begin{figure}[!h]
\noindent
\centering
\includegraphics[width=0.5\textwidth]{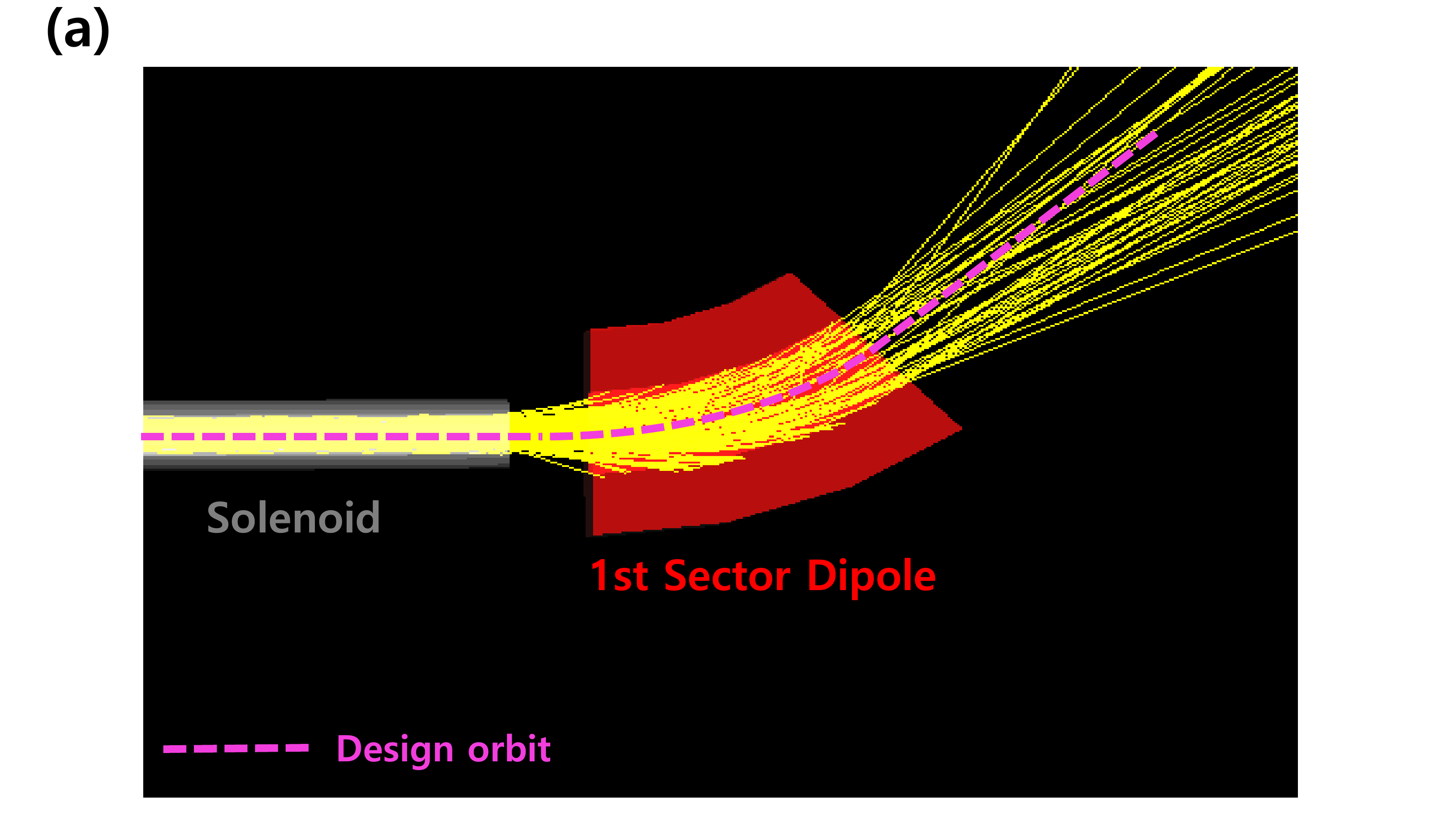}
\hspace{-0.4cm}
\includegraphics[width=0.5\textwidth]{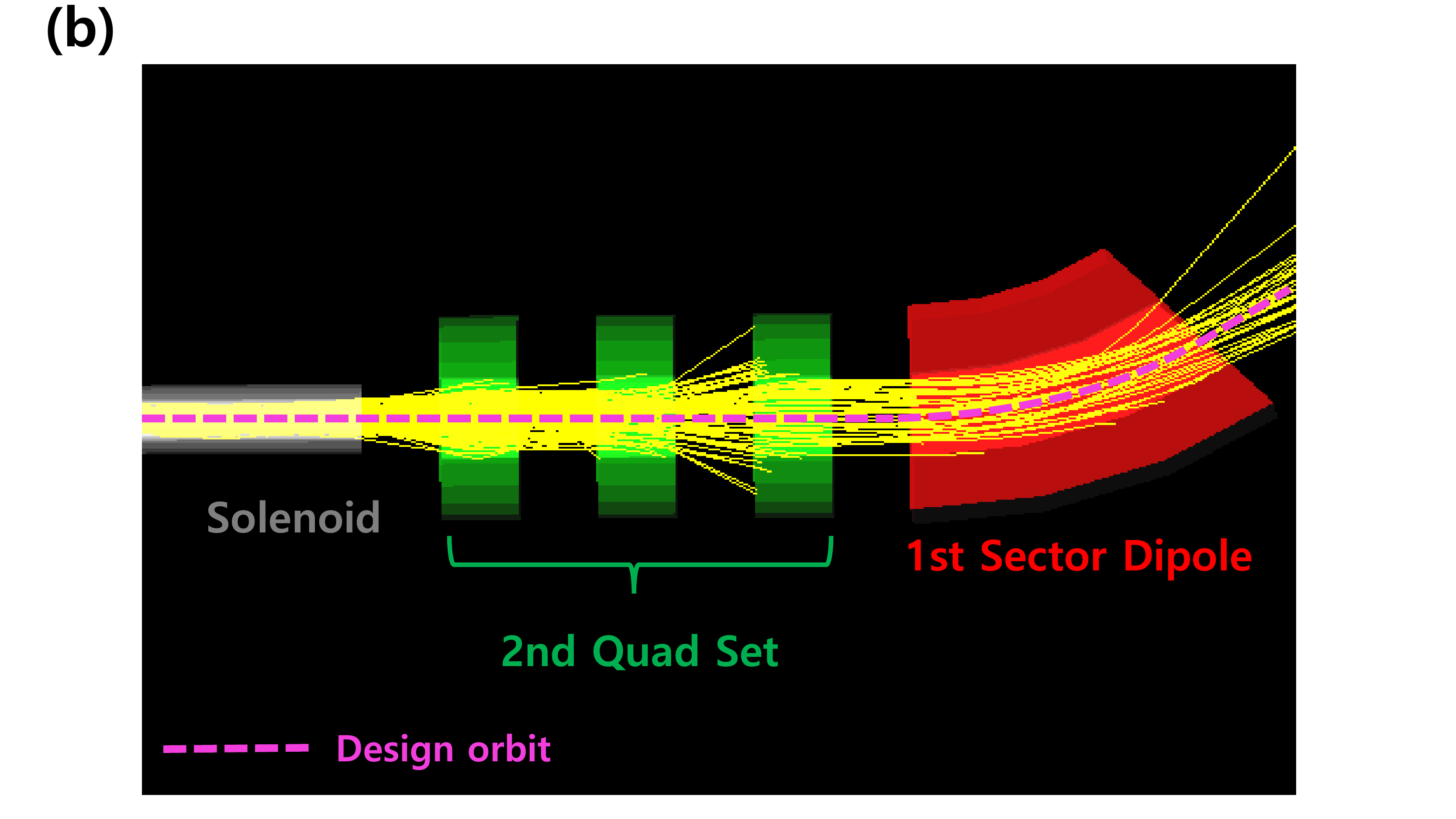}
\caption{\label{Eff_Q2_G4blgui}The effect of the second quadrupole set. (a): A lot of antimuons(yellow line) collide into the dipole's body. (b): Smaller antimuons collide to the dipole's body.}
\par
\end{figure}

However, the antimuons exiting the solenoid have a wide range of transverse momenta since they have just finished the helical motion. In this situation, the antimuons can not be bent ideally because the large transverse momentum forces the antimuons to collide into the dipole's body as shown in FIG. \ref{Eff_Q2_G4blgui}-(a). Hence, we used the second quadrupole set before the sector dipole to focus the antimuons. The field gradients of the second quadrupole set to focus about 130 MeV/c antimuons are given as 3.70 T/m, -4.08 T/m and 2.7 T/m, respectively and these successfully transport the antimuons to the first sector dipole with low transverse momenta as shown in \ref{Eff_Q2_G4blgui}-(b).

\begin{figure}[!h]
\noindent
\centering
\vspace{-4mm}
\includegraphics[width=0.5\textwidth]{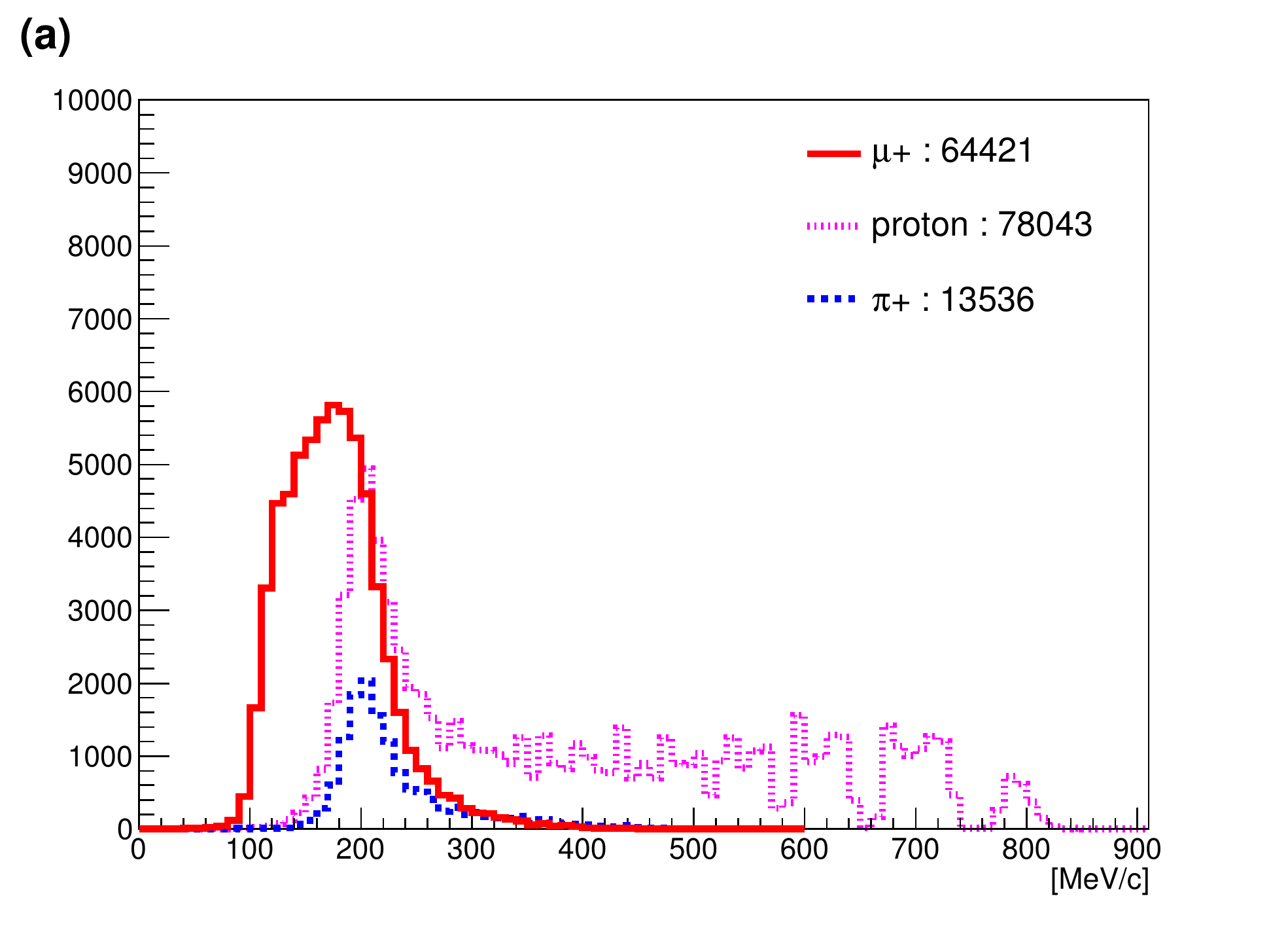}
\hspace{-0.5cm}
\includegraphics[width=0.5\textwidth]{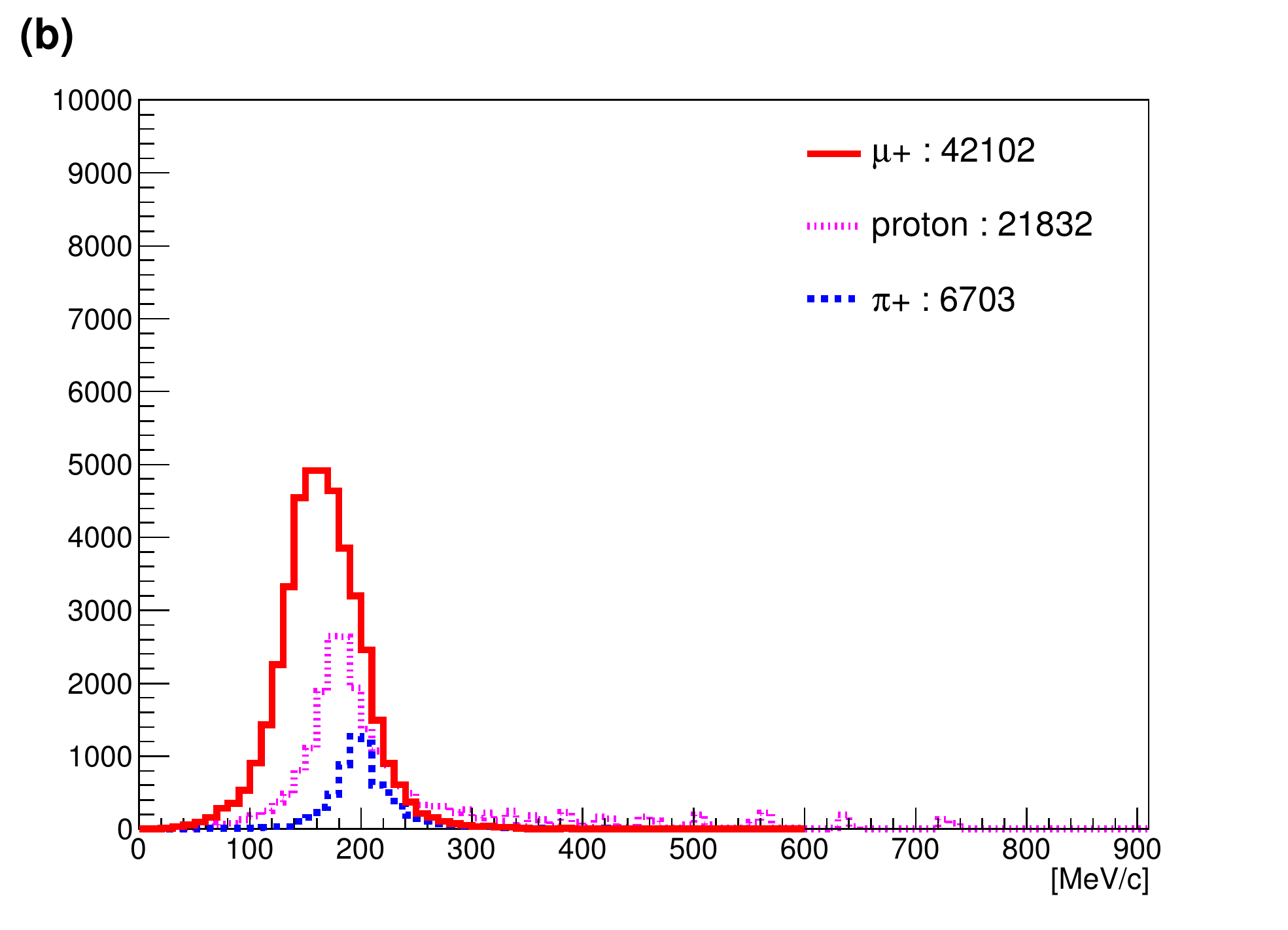}
\caption{\label{Eff_1st_SD1}The momenta distribution of the protons, pions, and the antimuons ($5\times10^{9}$ protons on the fixed target). (a): Entering the first sector dipole. (b): Exiting the first sector dipole The first sector dipole removes about 70 \% protons, 50 \% pions and 34 \% antimuons.}
\par
\end{figure}

The first sector dipole applies 0.26 T in the -$\hat{y}'$ direction and its angle is $40^{\circ}$. The 130 MeV/c antimuons on the design orbit follow an arc with 1.67 m radius in the first sector dipole and are ideally bent without the collision. Also, $40^{\circ}$ forces the positively charged particles whose momentum range is below 80 MeV/c or above 200 MeV/c to collide with the dipole's body. The effect of the first sector dipole is shown in FIG. \ref{Eff_1st_SD1}.

\begin{figure}[!h]
\noindent
\centering
\vspace{-4mm}
\includegraphics[width=0.5\textwidth]{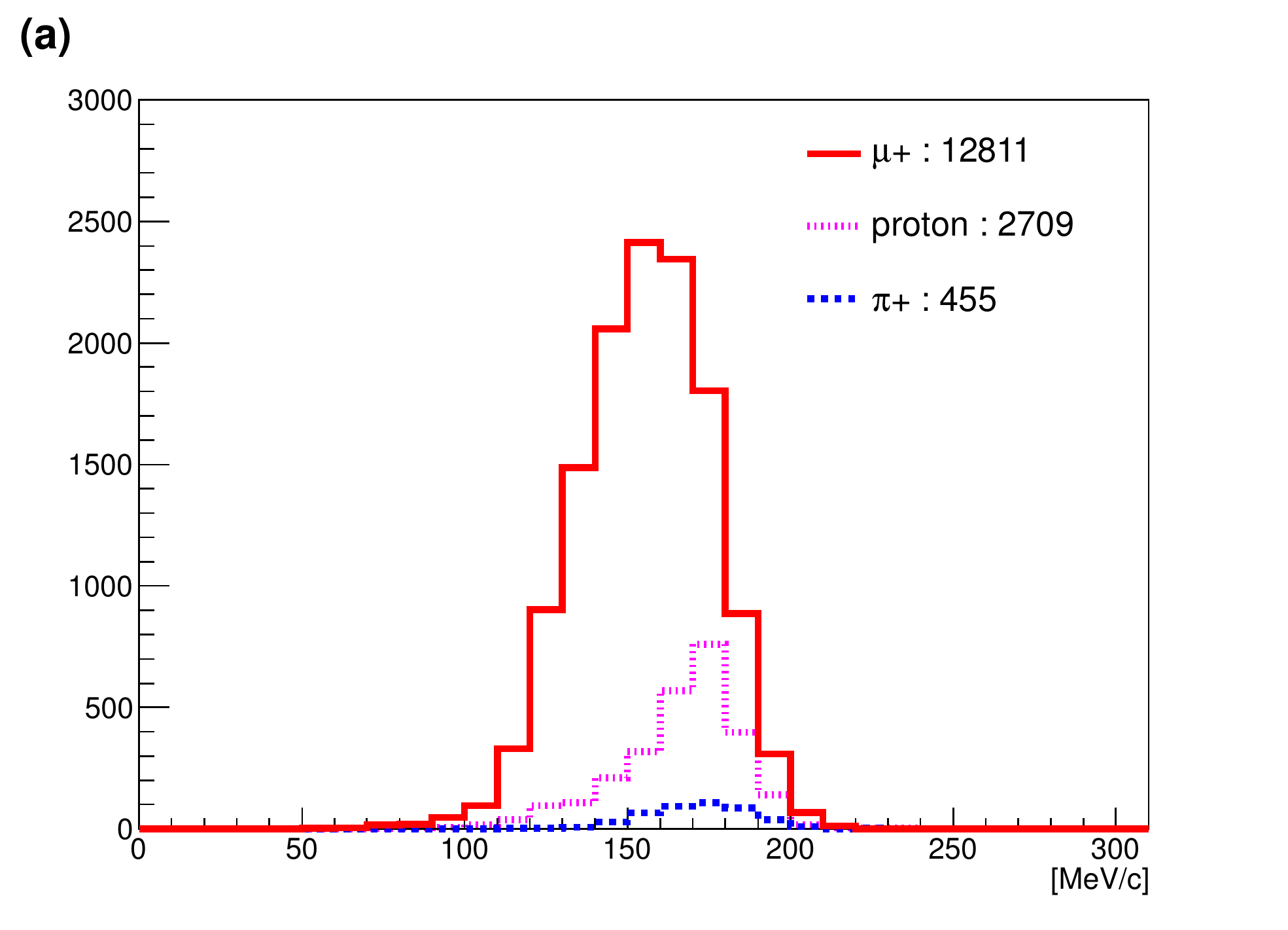}
\hspace{-0.5cm}
\includegraphics[width=0.5\textwidth]{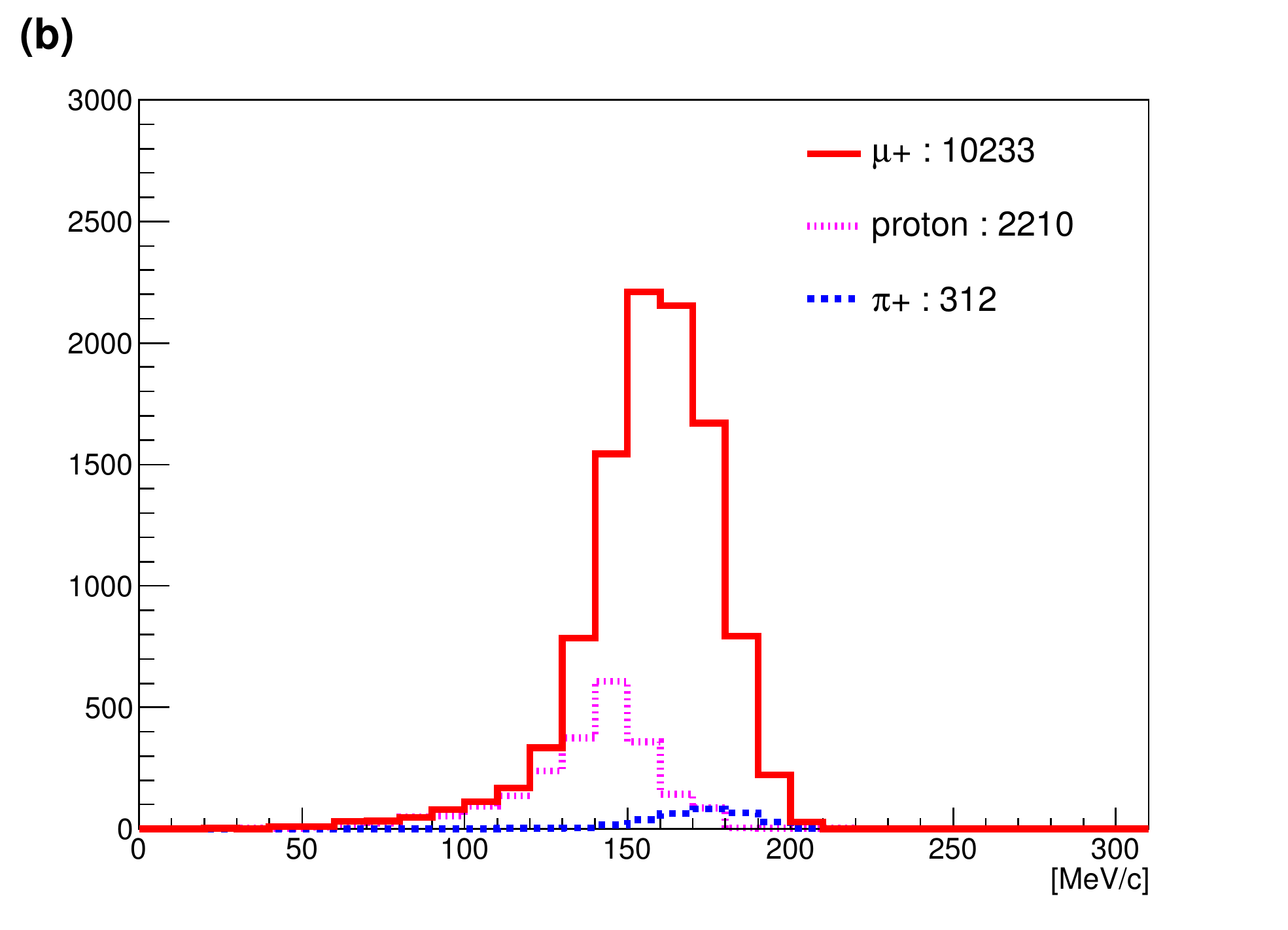}
\caption{\label{Eff_2nd_SD}The momenta distribution of the protons, pions, and the antimuons ($5\times10^{9}$ protons on the fixed target). (a): Entering the second sector dipole. (b): Exiting the second sector dipole: About 18\% protons, 31\% pions and 20\% antimuons are removed by the second sector dipole.}
\par
\end{figure}

The antimuons exiting the first sector dipole have a wide range of transverse momenta since not all the antimuons bend along the design orbit. Also, the antimuons still contain much contamination as shown in FIG. \ref{Eff_1st_SD1}-(b). Hence, we applied the focusing and momentum selection once more by using the third quadrupole set and second sector dipole successively. The field gradients of the third quadrupole set to focus about 130 MeV/c antimuons are given as 4.37 T/m, -4.04 T/m and 2.40 T/m, respectively and the second sector dipole removes about 18\% protons and 31\% pions as shown in FIG. \ref{Eff_2nd_SD}.

\begin{figure}[!h]
\noindent
\centering
\vspace{-4mm}
\includegraphics[width=1\textwidth]{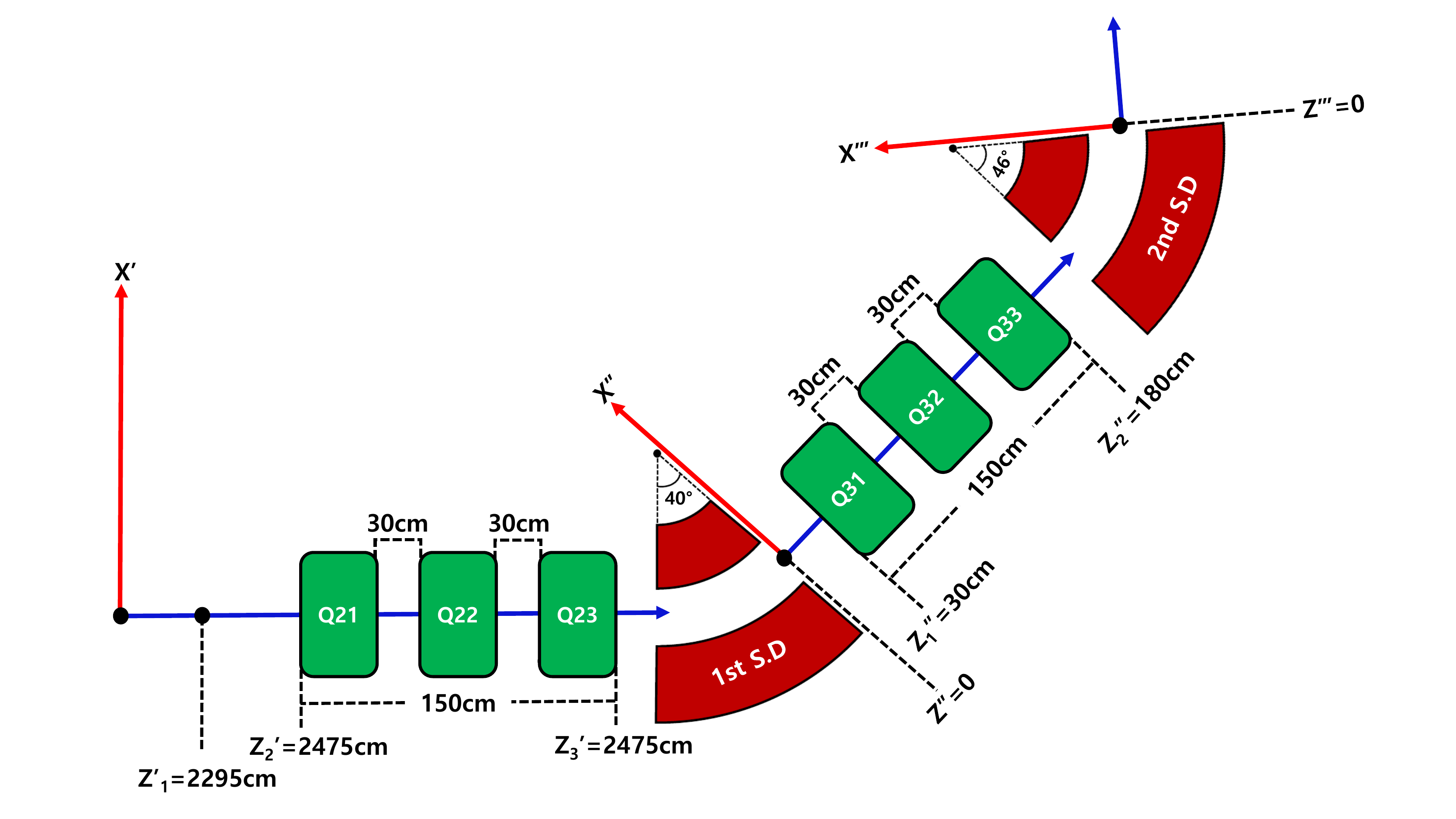}
\caption{\label{Front_Muon_Collecting}The front part of the antimuon collecting and purification section.}
\par
\end{figure}

\begin{figure}[!h]
\noindent
\centering
\vspace{-5mm}
\includegraphics[width=0.9\textwidth]{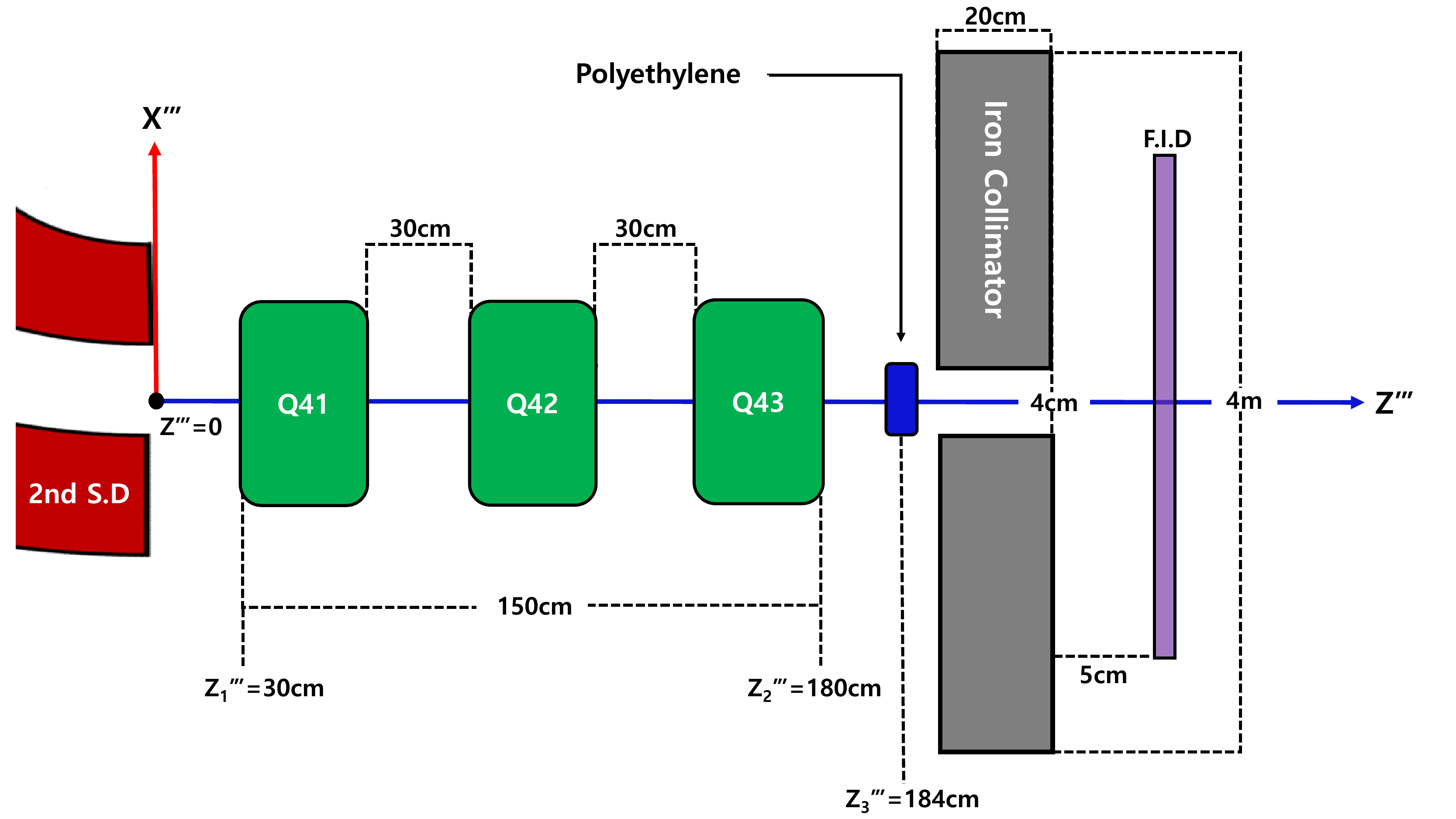}

\caption{\label{Last_Before}The final part of the antimuon collecting and purification section. The label F.I.D means the final imaginary detector that consist of vacuum. It is a circle that is of 80 cm radius and 1 mm length.}
\par
\end{figure}

The front part of the antimuon collection section is shown in FIG. \ref{Front_Muon_Collecting}. From now on, we do not apply the momentum selection any more since the momentum range of the protons and the pions overlap with the momentum range of the antimuons. We should use another way, a proton absorber. We assumed the maximum momenta of the protons exiting the second sector dipole is 200 MeV/c and used the polyethylene as the proton absorber. The non-relativistic Bethe-Bloch equation tells the minimum length of the polyethylene to stop 200 MeV/c protons should be above 0.7 cm. Hence, we placed polyethylene whose length is about 2 cm. This filtering process is carried out with the focusing process by using the 4th quadrupole set as shown in FIG. \ref{Last_Before}. We focused the antimuons whose momentum range is around 130 Mev/c after the second sector dipole and the field gradients of the 4th quadrupole set are given as -3.0 T/m, 4.73 T/m and -6.51 T/m, respectively. All of the protons are removed in this process.

\begin{figure}[!h]
\noindent
\centering
\vspace{-4mm}
\includegraphics[width=0.5\textwidth]{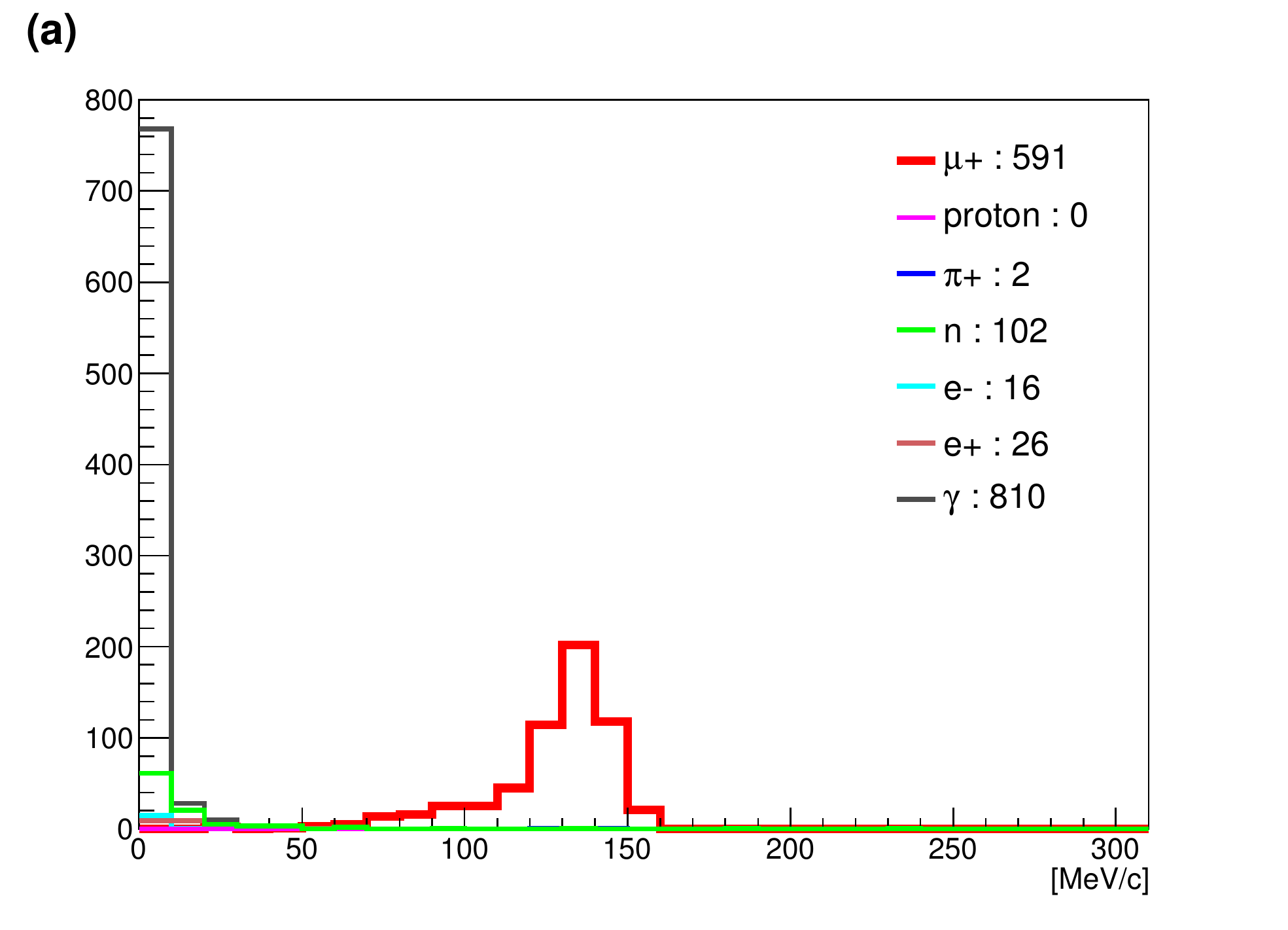}
\hspace{-0.5cm}
\includegraphics[width=0.5\textwidth]{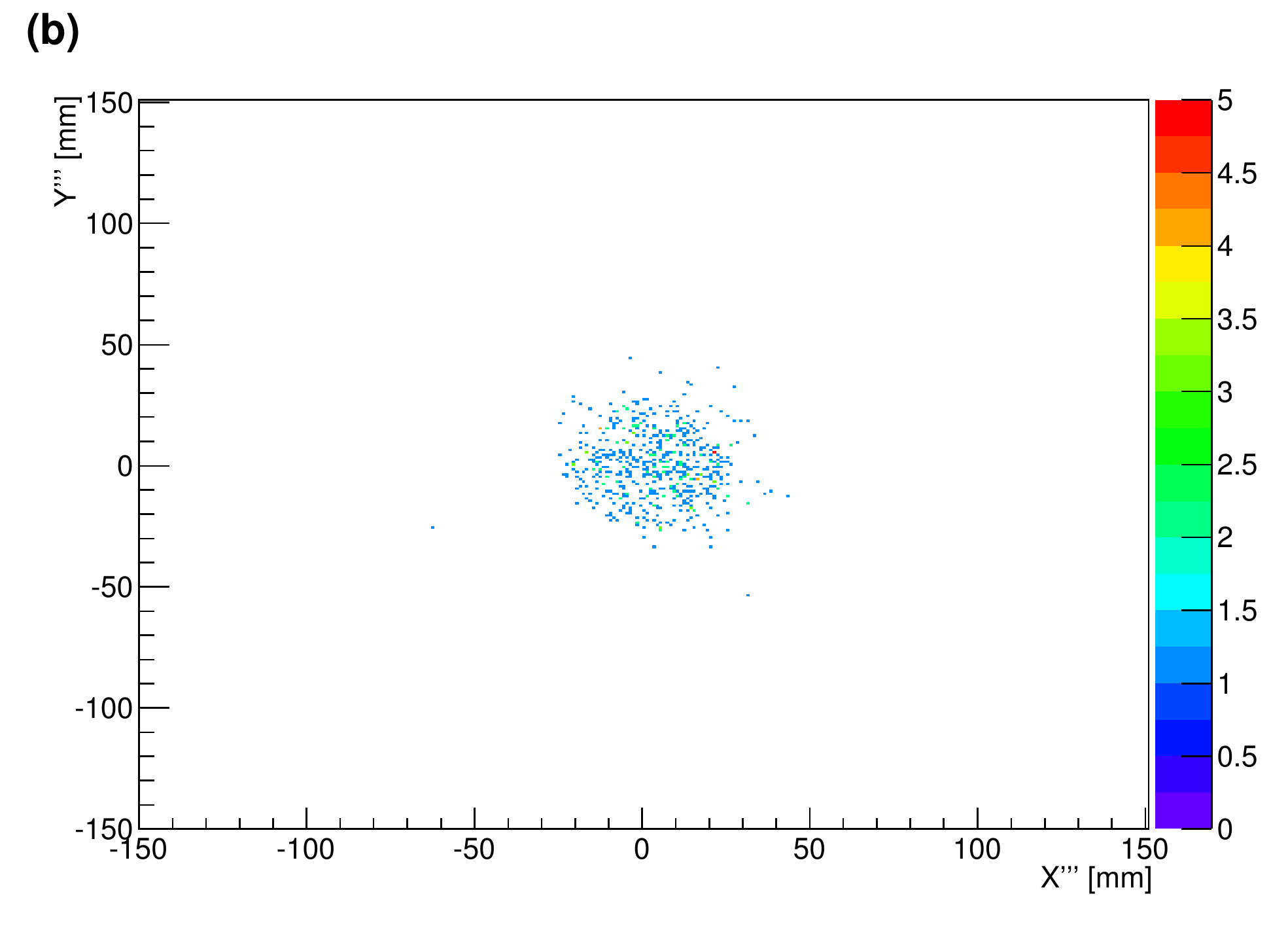}
\vspace{3mm}
\caption{\label{Last_Beam}The final beam information at the final imaginary detector ($10^{10}$ protons on the fixed target) (a): The momentum distribution of the antimuons and the contamination. (b): The position of the antimuons: Most of them exist in the circle region whose radius is 3 cm.}
\par
\end{figure}

Finally, we used the collimator to cut off the beam except for the center region. We assumed the maximum momenta of the antimuons and $\pi^{+}$ are 200 MeV/c and considered the both the minimum ionizing particles. Bethe-Bloch equation tells the enough length of the iron collimator should be above 11 cm \cite{Data_Book}. Hence, we used the sufficiently heavy iron collimator whose inner radius is 2 cm, outer radius is 2 m and the length is 20 cm. and obtained the centered antimuon beam with high purity as shown in FIG. \ref{Last_Beam}. Although we used the collimator whose inner radius 2 cm, the antimuons exist on the circle whose radius is about 3 cm by the effect of the transverse momenta.

\section{CONCLUSIONS}

We have designed the optimized muon beamline can transport 130 MeV/c$\sim$140 MeV/c antimuons dominantly. RAON is expected to provide 600 MeV kinetic energy proton beam with the beam current of 660 $\mu$A and this means the number of protons that enter the graphite target will be about 4$\times$$10^{15}$ s$^{-1}$. Hence, the optimized muon beamline is expected to give about $2.4\times10^{8}$ antimuons per second to a circle of 3 cm radius. In terms of muon rate, this is competitive with world leading muon beam facilities \cite{PSI_rate} ~\cite{jparc_rate}.

\section{Acknowledgements}
This research was supported by Basic Science Research Program through the National Research Foundation of Korea(NRF) funded by the Ministry of Science, ICT and Future Planning(2011-0016554) and  funded by the Ministry of Education(No. 2013R1A1A2062722). Also, it was supported by a Korea University Grant.

\end{document}